\documentclass[12pt,preprint]{aastex}

\shorttitle{}
\shortauthors{Newsham \& Terndrup}

\begin{document}

\title{Observational consequences of the hypothesized helium rich
stellar population in $\omega$ Centauri.}

\author{G. Newsham and D. M. Terndrup}
\affil{Department of Astronomy, The Ohio State University,
    Columbus, OH 43210-1173}
\email{newshamg@astronomy.ohio-state.edu,terndrup@astronomy.ohio-state.edu}

\begin{abstract}

In response to the proposed high helium content stars as an
explanation for the double main sequence observed in $\omega$
Centauri, we investigated the consequences of such stars elsewhere
on the color-magnitude diagram. We concentrated on the horizontal branch 
where the effects of high helium are expected to show themselves more
clearly. In the process, we developed a procedure for comparing
the mass loss suffered by differing stellar populations in a
physically motivated manner. High helium stars in
the numbers proposed seem absent from the horizontal branch of
$\omega$ Centauri unless their mass loss history is very different
from that of the majority metal-poor stars. It is possible
to generate a double main sequence with existing $\omega$ Centauri
stars via accretion of helium rich pollution consistent with
the latest AGB ejecta theoretical yields, and
such polluted stars are consistent with the observed HB
morphology of $\omega$ Centauri. Polluted models are
consistent with observed merging of the main sequences as opposed
to our models of helium rich stars. Using the
$(B-R)/(B+V+R)$ statistic, we find that the high helium bMS stars
require an age difference compared to the rMS stars that is too
great, whereas the pollution scenario stars have no such conflict
for inferred $\omega$ Centauri mass losses.

\end{abstract}


\section{Introduction}

The globular cluster $\omega$ Centauri has long been known to
exhibit a wide range of metallicities
\citep[e.g.,][]{dic67,fre75,sun96}, possibly in discrete
subpopulations. \citet{nor96} presented an extensive survey of
Ca-triplet abundances, revealing a majority metal-poor component
at ${\rm [Ca/H]} = -1.4$, an intermediate metal-poor peak at ${\rm
[Ca/H]} = -0.9$, and a long tail extending up to a ${\rm [Ca/H]} =
-0.3$. \citet{pan00} discussed wide-field $BI$ photometry and the
Norris et al.\ abundances, identifying four peaks at ${\rm [Ca/H]}
= -1.4$, $-1.0$, $-0.5$ and $-0.1$.  The [Ca/Fe] ratio is almost
flat as a function of [Fe/H] \citep{nor95} with an average value
$\langle{\rm [Ca/Fe]}\rangle \approx +0.4$, so the first two peaks
of the abundance distribution have ${\rm [Fe/H]} \approx -1.7$ and
$-1.3$, respectively.

The peculiarities of $\omega$ Centauri were further revealed by
the discovery of multiple turn-offs and a double main sequence
(MS) in HST photometry by \citet{bed04}. As seen in their Figure
1, the two main sequences are clearly separated by $\Delta (V-I)
\sim 0.06$ over several magnitudes in $V$, with the region
between them almost devoid of stars. It can be seen that towards 
faint magnitudes the two main sequences merge together, this is even
more obvious in the latest data of \citet{vil07}. The blue main sequence (bMS)
is less populated than the red main sequence (rMS), comprising
about 25\% to 35\% of MS stars. As discussed in \citet{bed04}, if
metallicity alone were controlling the morphology of the main
sequence, the MS would be about 0.03 magnitude in width with a
concentration toward the blue edge corresponding to the ${\rm
[Fe/H]} = -1.7$ majority ($\approx 65\%$) population, a tail towards
the red of stars corresponding to the intermediate metallicity
population ($\approx 30\%$) and a small ($\approx 5\%$) of the
stars redder still.

Of the various hypotheses put forth in \citet{bed04} to explain
the double main sequence, most intriguing has been the idea that
the bMS stars have an unusually high helium content. \citet{nor04}
noted that the rMS and bMS stars were present in a ratio of 2:1,
which was approximately like the ratio of stars with ${\rm [Fe/H]}
= -1.7$ to those near ${\rm [Fe/H]} = -1.3$. Taking the Revised
Yale Isochrones \citep{gre87} and fitting them to a synthetic
color magnitude diagram (CMD) of $\omega$ Centauri, Norris found
that the double main sequence can be reproduced if the
intermediate metallicity population is more helium rich by $\Delta
Y \approx 0.10 \-- 0.15$ than the metal-poor population.

\citet{pio05} undertook a spectroscopic investigation of the bMS
and rMS stars in $\omega$ Centauri. They obtained VLT spectra of
17 rMS stars and 17 bMS stars which were combined respectively to
form a single rMS and single bMS composite spectrum. They found
that the rMS stars have ${\rm [Fe/H]} = -1.57$ and the bMS stars
have ${\rm [Fe/H]} = -1.26$, close to the largest two peaks in the
distribution of abundances on the giant branch. \citet{pio05} also
fit new stellar isochrones to the bMS and showed that it can
best be modelled with 0.35 $\lesssim$ Y $\lesssim$ 0.45, the best
value being $Y = 0.38$. This confirms the
\citet{nor04} result but with metallicities directly from the MS
and more up-to-date stellar models. Similar conclusions were drawn
by \citet{lee05}, who determined a best fit to the observations
implied a bMS helium content of 0.38, implying $\Delta Y \approx
0.15$. \citet{dan05} have reached a similar conclusion of a spread
of helium content for the main sequence stars of NGC 2808 using
HST photometry. They noted that the color distribution is not
Gaussian and is wider than that expected for a single metallicity
population. They found some 20\% of the stars are much bluer than
expected and from their stellar models conclude that the helium
mass fraction of these stars is $Y \sim 0.4$.

Attempts to explain the high helium content of the bMS stars have
not been entirely successful. If massive stars in the metal-poor
population were responsible for both the helium and metal
enrichment of the intermediate metallicity stars, values of
$\Delta Y / \Delta Z \sim 100$ are implied \citep[e.g.,][]{nor04},
far in excess of the canonical value $3 - 4$ \citep[e.g.,][]{pag92}.
\citet{pio05} noted that massive stars should also produce a large
amount of carbon, nitrogen, and oxygen, but in $\omega$ Centauri
the total $[(\rm{C+N+O)/Fe}]$ abundances for all stars is about +0.4 dex
\citep{nor95}.  This is also the case for enrichment by AGB star
winds \citep{kar06}. \citet{bek06} extensively discussed
enrichment by intermediate mass AGB stars, massive stars that produce
helium-rich winds \citep[see][]{mae06} and SNe II. They
concluded that for reasonable Initial Mass Function (IMF)
choices $\omega$ Centauri could not have survived disintegration
due to cluster mass loss. The most serious problem with the AGB
scenario is that the total mass of ejecta from rMS AGB stars is
too small to be consistent with the observed fraction of the bMS
of $\omega$ Centauri. If $\omega$ Centauri formed with a very
unusual IMF, producing little or no SNe II, then the observed bMS
fraction can be reproduced; that is, however, inconsistent with the
known presence of neutron stars in $\omega$ Centauri. Similarly,
the number of massive stars with helium-rich winds do not
contribute a large enough mass fraction of ejecta for allowable
IMF choices. In addition, there would be little difference in
metallicity between the rMS stars and the massive star wind
ejecta, which is at odds with the \citet{pio05} result on the
metallicity difference between the bMS and rMS populations.
Conversely, the problem with the SNe II scenario is that the heavy
element abundances would be too high to be consistent with the bMS
stars forming from SNe II ejecta. The conclusion was that neither
alone or combined could these three sources explain the bMS stars
as being formed exclusively from the ejecta of rMS stars. Recent work
by \citet{cho07} have concluded the only way to achieve such high
helium enrichments without a commensurate level of metal production
is by appealing to Population III star ejecta as the source of the
material that formed the bMS. However, they admit this scenario is itself
of an extreme nature.

Several studies have tried to find the evolutionary descendants of
the He-rich population on the horizontal branch (HB).  Stars with
high ($Y \gtrsim$ 0.35) helium content would reach the tip of the
RHB (TRGB) with a reduced mass since they have shorter lifetimes.
This would make the HB both hotter (bluer) and more luminous when
compared to stars of lesser helium content \citep{dan02}.
\citet{dcr00} \& \citet{mom04} noted that $\omega$ Centauri does
have a noticeably blue HB with substantial numbers of Extreme
Horizontal Branch (EHB) stars with an effective temperature
($T_{\rm eff}$) greater than 20,000K. This has been noted as
qualitative evidence of the bMS stars progeny populating the HB of
$\omega$ Centauri \citep{nor04,pio05,lee05}.

Recent observations of the RR Lyrae variables in $\omega$ Centauri
by \citet{sol06} showed spectroscopically the presence of both
metal-poor and intermediate metallicity stars in their sample of
74 RR Lyrae variables. These intermediate metallicity variables
are not consistent with a high helium enhancement, in that they do
not have a higher luminosity and they have an ordinary
relationship between period and luminosity. Thus, at least some
intermediate metallicity stars have normal helium abundances. As
\citet{sol06} themselves noted, this is not necessary a strong
argument against the enhanced helium hypothesis, since bMS stars
would not appear in the RR Lyrae instability strip unless they are
some 4 Gyr younger than the rMS population or have undergone a
radically different mass loss history.

In this paper we perform a quantitative calculation of the color
distribution of HB stars for various helium abundances.  We also
investigate whether accretion of helium-rich material onto
existing stars can produce the observed double main sequence, and
follow the evolution of these stars to the HB. A description of
the stellar evolutionary calculations we performed is presented in
\S  2, which also discusses models of stars with homogeneous high
helium abundance.  Models with a high surface abundance of helium
are presented in \S 3. In \S 4 we summarize our results.

\section{Models with homogeneous enhanced helium}

\subsection{Stellar Evolution Models}

We computed evolutionary tracks appropriate for $\omega$ Centauri
using the modern version \citep{sil00} of the Yale Rotating
Stellar Evolution Code \citep[YREC,][]{gue92}; in the
computations, the rotation routines were turned off. Nuclear
reaction rates are taken from \citet{gru98} and the heavy element
mixture is that of \citet{gre93}. Gravitational settling of helium
and heavy elements is included in these models, as in
\citet{bah90} and \citet{tho94}.  We also include neutrino losses
from photo, pair, bremmstrahlung and plasma neutrinos, following
\citet{ito96}.

We use the OPAL opacities \citep{igl96} for temperatures $\log T
\geq 4$. For lower temperatures, we employ the molecular opacities
of \citet{ale94}. The alpha-enhanced low {\it T} opacities
are known to be in error \citep{wei06} but this does not
affect our calculations as we used the solar mixture opacities.
For regions of the star with $\log T \geq 3.7$,
we use the 2001 version of the OPAL equation of state
\citep{rog96}\footnote{Web updates at http://www-phys.llnl.gov/Research/OPAL/Download}
and for $\log T < 3.7$ we take the equation of state from \citet{sau95}.

For the surface boundary conditions, we use the stellar atmosphere
models of \citet{all95}, which include molecular effects and are
therefore appropriate for low mass stars. We use the standard
B\"{o}hm-Vitense mixing length theory \citep{boh58,cox68} with $\alpha
= 2.013$, obtained by calibrating a solar model against
observations of the solar radius ($6.9598 \times 10^{10}$ cm) and
luminosity ($3.8515 \times 10^{33}$ erg s$^{-1}$) at the present
age (4.57 Gyr) and metal fraction ($Z = 0.01757$) of the Sun. By
comparison, models of the Sun excluding diffusion require $\alpha
\approx 1.7$ \citep[e.g.,][]{pin03}.

The helium content of ``ordinary'' (i.e., not enhanced in helium)
$\omega$ Centauri stars was computed using the primordial helium
abundance abundance $Y_p = 0.23$ and the enrichment parameter
$\Delta Y / \Delta Z = 2.0$, in line with the treatment in the
Yale-Yonsei isochrones of \citet{yi03}.  The value for $Y_p$ is
within, but on the low end of, the range of current estimates
\citep[e.g.,][and references therein]{bono02,thuan02,oli04}
although some values are considerably higher
\citep[e.g.,][]{fk06}. This would yield a helium abundance for the
Sun of $Y = 0.265$ in models lacking diffusion. Models compatible
with helioseismology which include both rotational mixing and
diffusion have a surface abundances $Y_{\odot,{\rm surf}} = 0.249
\pm 0.003$ and an initial solar composition $Y_\odot = 0.274$,
$Z_\odot = 0.019$ \citep{bah01}. Such values would imply $\Delta Y
/ \Delta Z = 2.3$, close to $\Delta Y / \Delta Z = 2.1 \pm 0.4$
inferred for nearby field stars \citep{jim03}, but larger than
$\Delta Y / \Delta Z = 1.3 \pm 0.2$ from open clusters in the
solar neighborhood \citep{an07}. At the low metallicities of the
stars in $\omega$ Centauri, however, the properties of the models
for unenhanced stars are insensitive to the exact value of the
helium enrichment parameter.

All models began with a zero-age main sequence (ZAMS) model
evolved from the deuterium-burning birthline.  The ZAMS was
defined as the point at which the core hydrogen abundance had
dropped by 2\% from the initial value.  Horizontal branch models
were created by rescaling the core masses, metallicity and
envelope masses of horizontal branch models created for the YREC
code provided by A.\ Sills (private communication). The core
masses were determined from the corresponding RGB tip (TRGB) core
masses from our evolved models and Zero Age Horizontal Branch
(ZAHB) sequences developed by rescaling the envelope masses to
define the ZAHB. This method of creating ZAHB sequences has been
found to be adequate by \citet{ser05} with deviations at the few
percent level being found for the hottest HB stars when compared
to ZAHB models calculated directly through the core helium flash.

We calculated theoretical isochrones by interpolating the model
evolutionary tracks using a scheme based upon the algorithm of
\citet{ber92}. The tracks, all with ages above 7 Gyr, have a
simple topology, so we used five major evolutionary points to
create the isochrones. These were the zero age main sequence
point, the point of the exhaustion of core hydrogen, the base of
the red giant branch, the red giant bump and the tip of the red
giant branch (i.e., at the helium flash). Observational colors and
magnitudes were calculated from $M_{\rm bol}$, log {\it g} and $T_{\rm
eff}$ using the transformations of \citet{van03}.

\subsection{Model populations}

\citet{nor04} and \citet{lee05} both modelled the bMS of $\omega$
Centauri as a population with ${\rm [M/H]} = -1.27$ and a helium
enhancement of $\Delta Y = 0.12 - 0.18$ over the rMS.
\citet{nor04} associated the bMS population with the second most
common RGB metallicity, while \citet{lee05} used the the bMS
metallicity as determined spectroscopically in \citet{pio05}.

In this present work we created isochrones for the two main
sequence populations using parameters shown in Table
\ref{parameters}, where we list the metal content, helium content,
metallicity, the TRGB total mass, and the TRGB helium core mass.
Here we are using models for the bMS that are homogeneous in helium
content; below we compute models in which the outer convective
zones are polluted with helium-rich material. Hereafter, the
proposed helium-rich population in all evolutionary states shall
be referred to as the bMS population, while the majority
metal-poor stars shall be called the rMS
population. We later discuss polluted models where we refer to a
population with the metallicity of the bMS stars but with a normal
helium content; we shall call this the intermediate metal poor
population (MintP). Its characteristics are also shown in Table
\ref{parameters}.

\begin{figure}
 \epsscale{0.85}
 \plotone{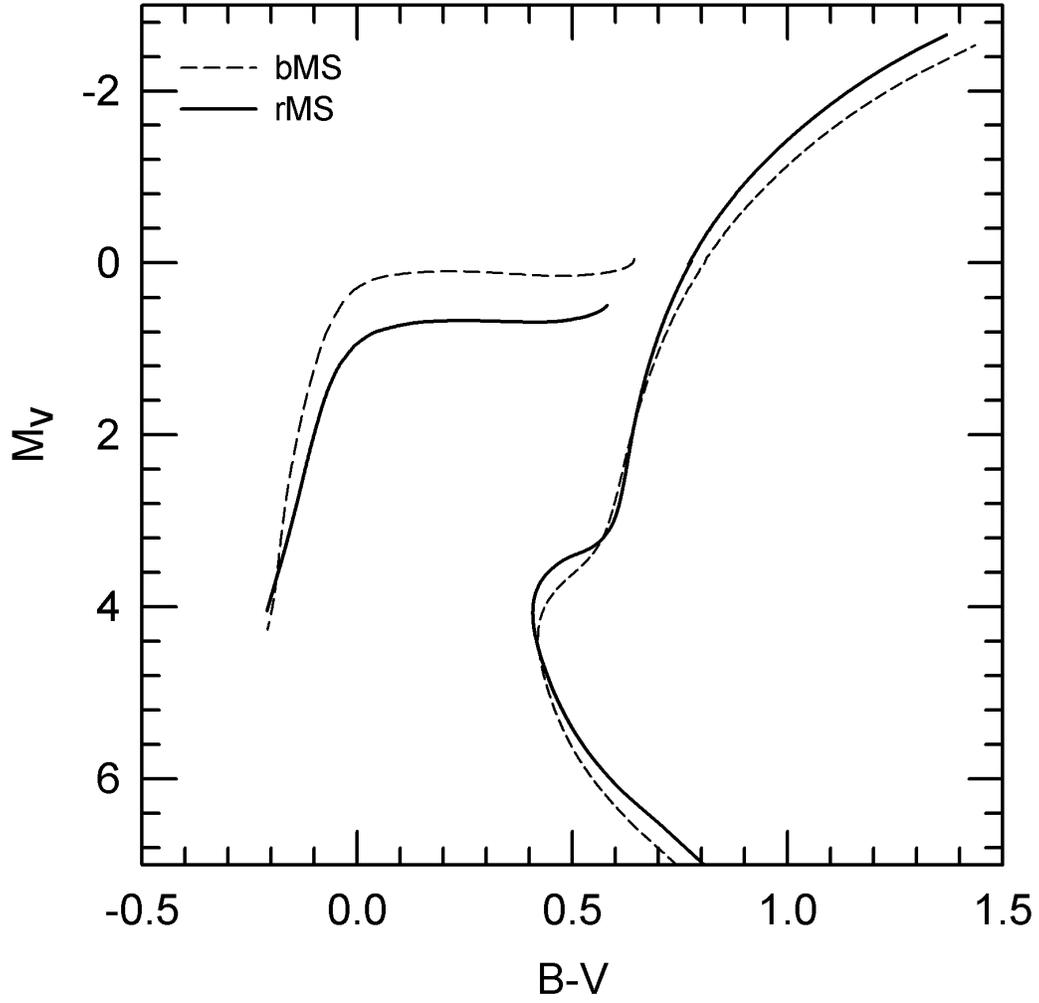}
\caption{The rMS and bMS 13 Gyr isochrones using the parameters in
Table \ref{parameters}.\label{fig:isochrones}}
\end{figure}

In Figure \ref{fig:isochrones} we show isochrones for the bMS and
rMS populations in the $M_{V}, \bv$ plane;  both groups have an
age of 13 Gyr. The two populations overlap at the turnoff (TO) but
separate again high on the RGB, where the bMS stars now appear on
the cool side of the rMS giant branch. As expected from theory,
the ZAHB of the bMS is more luminous (by $\sim 0.8$ magnitudes)
than that of the rMS.  Both sequences merge on the hot tail of the
blue HB.

\subsection{Mass loss methods}

In order to populate the ZAHB sequences we need to parameterize
the mass loss at the TRGB.  There are two approaches that are used
to calculate a distribution of HB masses.  In the first, mass loss
is parameterized with mean value ($\Delta M$) and a dispersion
$\sigma_{M}$;  the form of the mass loss distribution is often
taken as a Gaussian.  Using this approach, \citet{lee94} and
\citet{cat00} studied the HB color distribution of several
globular clusters, and found mean HB masses of $0.63 - 0.75
M_{\sun}$, implying mass losses of the order $0.1 - 0.2 M_{\sun}$.
The presence of extremely blue HB stars in some clusters implies
even greater mass losses, of the order $0.3 M_{\sun}$ (cf.\ Table
\ref{parameters}), since these stars have a very thin ($M \leq
0.002 M_\sun$) hydrogen envelope surrounding the helium core.

Alternatively, some authors calculate mass loss during the RGB
evolution, employing various empirical approximations.  The mass
loss scales with one or more physical parameters of the RGB
star \citep[see][for a summary]{cat00}. It has long been
questioned whether or not these formulae accurately estimate the
mass loss rates of RGB stars.  \citet{ori02} showed that the
observed loss rates are at least an order of magnitude greater
than the predictions of the empirical formulae, and the rates do
not seem to follow the dependence on luminosity, gravity, radius,
or metallicity which appear in the empirical expressions.  They
also concluded that the mass loss occurs only in the last $10^{6}$
years of RGB evolution (i.e., very near the TRGB) and is episodic
in nature; the observed time scale for mass loss is greater than a
few decades and less than a million years.

In this paper, we follow the first approach.  The models are run
to the TRGB with no mass loss.  At the helium flash, we generate a
distribution of HB masses to populate the ZAHB.  By doing so, we
do not need to follow the effect of mass loss on the structural
evolution of stars near the TRGB.

Because the rMS and bMS stars reach the TRGB with quite different
masses (Table \ref{parameters}), they may have a different average
mass loss.  Simply assuming a mean mass loss of order
$0.2M_{\sun}$ for the bMS would remove the entire envelope and
part of the helium core, yet we know from EHB star spectra that
hydrogen is present in these stars. We note that any mass loss
greater than $0.174M_{\sun}$ for the bMS stars at an age of 13 Gyr removes
the entire hydrogen envelope of the star.

We consider several cases for computing the mass loss for bMS
stars. The first case (Case I) corresponds to the situation in
which the mass loss takes place on time scales shorter than
the  Kelvin-Helmholtz time, perhaps as short as the bottom of the
range of time scales discussed by \citet{ori02}.  Whatever the
mechanisms operate to produce the mass loss, they have to remove
material from the star by doing work against the gravitational
potential. In Figure 2 we plot the acceleration of gravity at an
interior point against $\Delta M$;  the latter is the total
stellar mass minus the mass at the interior point.  Shown are TRGB
models of the bMS and rMS populations at an age of 13 Gyr. The bMS
model is distinctly more compact in structure than the rMS star.
Rapid mass loss, therefore, is likely to remove a smaller amount
of material in helium-rich stars.

The second case (Case II) assumes that the time scale for mass
loss is long, and that both populations have the same dependence of
loss rate on luminosity. This implies that the total mass loss
for the rMS and bMS stars would be in proportion to their
evolutionary timescales near the TRGB.  The bMS stars evolve more
rapidly than the rMS stars, so in this case they would also have a
lower amount of material removed, though more than in Case I.

We also compute (Case III) mass loss using an empirical mass loss
law, the Reimers prescription as generalized in \citet{cat00}.
Such a prescription assumes that the mass loss rate includes a
dependence on the luminosity, temperature and gravity of the star and that it
occurs over a large timescale. Thus the total mass loss can be
approximated as the product of the evolutionary timescale of the
star on the upper RGB and the mass loss rate from the Reimers
formula. In this case the mass loss is approximately the same in both rMS and bMS stars.

In Table \ref{deltam}, we show various $\Delta M$ for for rMS stars
and the corresponding cases for mass loss of bMS stars.   The
lowest relative mass loss is for Case I, while the highest is for
Case III.  In the discussion following we calculate HB models
using Case I. We justify this choice on the basis that it represents
the most conservative case of mass loss, the other cases would 
push the HB stars from the bMS population to even higher temperatures 
producing even more of a clump in the HB tail that is not seen. We also
feel that in light of the observational data of \citet{ori02} that
Case I is the physically most likely choice for mass loss. We 
implemented the mass loss by assuming
a $\Delta M$ with a small dispersion, $\sigma_{M}$, to be subtracted from
the TRGB mass for rMS stars. We found the mass of the immediate ZAHB star
progenitor from the mass of the star at the TRGB on the
appropriate isochrone. We took the $\Delta M$ value 
and found the local acceleration due to gravity, $g$, in the corresponding
rMS TRGB stellar model. We then applied these values of $g$ to
the bMS model and calculated the corresponding $\Delta M$ and
$\sigma_{M}$ for the bMS model.  The resulting ZAHB population
distributions were proportioned in accordance to the observed
numbers of rMS and bMS stars, with 70\% from the rMS and 30\% from
the bMS.

\subsection{Results}

Figure \ref{fig:hbmodels} displays an example of the relative
locations of the bMS and rMS stars on the HB.  The plot is in the
observational ($M_{V}, \bv$) plane.  The rMS population is shown
as open triangles, and the bMS population as filled circles. The
isochrones are also shown as solid and dashed lines, respectively. Here,
both populations have an age of 13 Gyr. The ZAHB is populated with
$\Delta M = 0.24 M_{\sun}$ and a dispersion of $\sigma_{M} = 0.015 M_{\sun}$ for
the rMS stars and the equivalent $\Delta M = 0.143 M_{\sun}$ 
with a dispersion of $\sigma_{M} = 0.009 M_{\sun}$ for
the bMS stars.  The ZAHB populations stay on the hot tails of both
ZAHB sequences, although the rMS population resides to the cooler
side and is redder by about 0.15 magnitudes in $\bv$ and brighter
on average by some 2 magnitudes in $V$.

In Figure 4 we see the effect of varying the mass loss
prescription with $\Delta M = 0.20$ and $0.28 M_{\sun}$ for
the rMS stars and equivalent $\Delta M$ of 0.118 and 0.168
$M_{\sun}$ for the bMS stars of 13 Gyr age. The stars are seen to
remain on the hot tail of the ZAHB sequence with $\bv$ $\lesssim
-0.05$. For the rMS population, stars of the lowest 0.20 M$_{\sun}$
mean mass loss start to appear on the horizontal region of the
ZAHB extending all the way up to $\bv \sim 0.2$.
 
The effect of age on the ZAHB distribution is also shown in Figure
4 for the bMS stars at 7 Gyr age ($\Delta M$ = 0.143 M$_{\sun}$).
This shows we can populate the ZAHB from the hot tail at $\bv$
$\lesssim -0.1$ at 13 Gyr all the way to $\bv$ = 0.6 at the
extreme red end for a 7 Gyr bMS population. To clearly differentiate 
a high helium bMS population we should look for a more luminous stellar
population at $\bv$ $\gtrsim$ 0.0 where the bMS stars populate a
ZAHB some 0.8 magnitudes brighter. This effect is likely to result
for bMS populations 2 to 6 Gyr younger than the rMS population
with greater mass losses requiring a corresponding greater age
difference. For bMS and rMS populations at the same 13 Gyr age the
differences on the ZAHB is primarily one of color alone and the
rMS stars being on average about 0.1 magnitudes in $\bv$ redder
than the bMS ZAHB stars, though for equivalent mass losses the
rMS stars are about $1 \-- 2$ magnitudes brighter in $V$.

\subsection{Comparison with observations}

In Figure \ref{fig:cmd}, we plot the  CMD of $\omega$ Centauri ground-based
data of \citet{rey04}.  Also shown are the 13 Gyr isochrones and
ZAHBs of the bMS and rMS populations from Figure
\ref{fig:isochrones}. We followed Rey et al.\ in adopting $E(\bv) =
0.12$ and $(m - M)_V = 14.1$.  Figure \ref{fig:rrstrip} shows the
HB only along with the location of the RR Lyrae instability strip
from \citet{bon95}.

On the CMD, the majority of the HB stars lie within $14.4 < V <
16.4$ and $0.00 < \bv < 0.32$. Following \citet{nor04}, we make
the assumption these are mainly from the rMS population.
If the age of the rMS population is 13 Gyr, then the rMS stars
have lost a mass of $0.168 - 0.247 M_{\sun}$. For a Case I mass
loss, the bMS stars at 13 Gyr would have lost $0.102 -
0.147M_{\sun}$.  This corresponds approximately to the location of
the filled circles on Figure \ref{fig:altdeltam}, so we should see
the bMS stars between $-0.2 \leq \bv \leq -0.15$ and on average about $1.5 \-- 3.0$
mag fainter in $V$ than the main HB.  This would place them on
the blue tail of the helium-rich HB, with $16.5 \leq V \leq 18.5$.
For mass loss amounts described by Case II or III, the bMS stars
would be even bluer and fainter forming a distinct clump at the
bottom of the HB tail. Although there are many such stars on the CMD (Fig.\
\ref{fig:cmd}) they are not present as expected from the relative
numbers on the main sequence (\citet{nor04} reached a similar conclusion) 
in which the bMS composes about 37\% of the bMS 
plus rMS total \citep[e.g.,][]{fer04,sol07,vil07}.

We also checked the deeper $\omega$ Centauri HST CMD of \citet{fer04} using data
furnished by A. Sollima (private communication).
In analyzing the HST data we find that the EHB stars compose
some $25\% \pm 3\%$ of the HB population due to the bMS and rMS stars.
This should be contrasted with the $37\% \pm 2\%$ derived in the latest MS data
by \citet{vil07}. The 10\% of stars of higher metallicity that 
are not attributed to the bMS and rMS populations would not appear
in the main clump of the HB for the inferred rMS mass loss as determined from our theoretical
models discussed above so we counted from the main clump and 
blueward giving us the rMS and bMS stars only.
Note that since the HB lifetime for both rMS and bMS stars is similar, as
expected from theory and confirmed in our models, then this difference 
in the number of bMS to rMS HB stars is not due to the effects of a 
different HB evolutionary lifetime.  Even if {\it all} 
the stars in the blue tail of the HB are in the bMS population, they number
only about 25\% of the stars on the main part of the HB.

It does not help to assume that the bMS stars are younger than 13
Gyr, and so have redder colors.  If the bMS population is 2 Gyr
younger, our models predict that they would be found on the ZAHB
with colors $0.05 < \bv < 0.16$ and brighter than the rMS HB
by 0.8 magnitudes in $V$. When we count the stars in this
region of the CMD, we find that they number $\sim 1.3\%$ of the
number of stars main clump on the HB.   Some of these stars in
this region of the CMD are undoubtedly rMS stars that have evolved
off the ZAHB, thus reducing the number of possible bMS stars in
that region. It thus appears that we do not see the substantial
number of stars in the CMD at the location we would expect the bMS
stars to appear. These calculations are also consistent with the
conclusions of \citet{sol06}, who failed to find any RR Lyrae
stars in $\omega$ Centauri that were helium rich, as indicated by
a high luminosity. Such stars would be in the instability strip
only if they were $2 \-- 5$ Gyr younger than the rMS population.  Such
young ages are excluded by the analysis of the turnoff region
\citep{sol05,lee05}.

Using the Rey et al.\ CMD, we also calculated the $(B-R)/(B+V+R)$
statistic \citep{lee94}, where $B$, $V$ and $R$ are the numbers of
red HB (RHB), RR Lyrae variables and blue HB (BHB) stars
respectively.  Following \citet{rey04}, we define the RHB stars as
having $14.2 < V < 14.85$ and $0.55 < \bv < 0.71$. In total we
find 2781 HB stars in the $\omega$ Centauri CMD of which 2497 are
BHB stars, 149 in the RR Lyrae instability strip, and 135 are RHB
stars. This gives us $(B-R)/(B+V+R) = 0.85 \pm 0.01$, with the
error arising from counting statistics. We shall use this result
later as a discriminator in discussing polluted stellar models
versus homogeneous helium-rich bMS stars. Using the HST data of
\citet{fer04} we calculated $(B-R)/(B+V+R) = 0.85 \pm 0.01$ which
is identical to the value we determined using the \citet{rey04} data.

\section{Pollution models}

\subsection{Construction of the models}

As noted by \citet{bek06}, the amount of helium required to form
a later generation of helium rich bMS stars is far greater than
would be produced by massive stars in the rMS population, unless
the initial mass function was top-heavy.  The large amount of
helium could have come from the ejecta of a much larger stellar
population, perhaps an ancient host galaxy of which the current
$\omega$ Centauri is the remaining nucleus;  this was the scenario
advanced by \citet{bek06}. Alternatively the helium could be
the end product of an initial Population III population as advanced 
by \citet{cho07}.

An alternative scenario, which we admit is somewhat contrived, is
that helium-rich material was accreted onto pre-existing rMS or
MintP stars;  the latter have $\rm[M/H] = -1.27$ but an ordinary
helium abundance. Such stars will appear on the hot side of the
rMS, as required by observation of the double main sequence, but
the amount of helium required would be far less than in the case
of bMS stars that are helium-rich throughout.  As the
helium-polluted stars ascend the giant branch, the deepening
convection zone would erase the abundance gradient in these stars,
and the final product would be expected to have TRGB total masses
and core masses that are more like those of the rMS population.

The question of accretion onto globular cluster stars has been addressed
in the past by \citet{tho02}. The mechanism 
for this is the classical Bondi accretion \citep{bon52} in which
the star, undergoing steady motion, accretes material 
in a spherically symmetric manner from an external medium.
As shown by the \citet{tho02} calculations, $\omega$ Centauri
is not considered to be a good example of an environment likely to allow much
accretion despite its deep gravitational well, though this may well have been
different in the past especially if $\omega$ Centauri was originally 
part of a larger system as is often proposed. This pollution scenario 
has also been considered by \citet{tsu07}.

The starting models for the pollution scenario investigation were
taken to be either the MintP or the rMS populations as shown in
Table \ref{parameters}. We created a set of models with
total fractional stellar mass accreted, composition and time delay
from the ZAMS, with values shown in Table 3. The accreted material
was of metal content $Z = 0.001$ so that the
polluted stellar models always maintained the surface metallicity
${\rm [M/H]} = -1.27$ of the observed bMS population main
sequence. These models ranged from $0.40 \-- 0.90 M_{\sun}$ in
0.05 $M_{\sun}$ increments. To each of these stars mass was
accreted as a fraction of the total starting mass of the stellar
model at a rate $10^{-10} M_{\sun}$ yr$^{-1}$, starting at a given
age of the stellar model ranging from the ZAMS point (age of 0.0)
up to a maximum of 9 Gyr in 3 Gyr increments. We chose our accretion
rate based upon the calculations of \citet{tho02} which show that
for typical globular cluster environments the accretion rates expected
are of the order $10^{-10} M_{\sun}$ yr$^{-1}$.

In Figure \ref{fig:modelseq} we show the 13 Gyr main sequence
isochrones for the rMS and bMS populations listed in Table
\ref{parameters}.  The accretion of helium-rich material will drag
the rMS isochrone to the left and up in the $M_{\rm bol}, T_{\rm eff}$
plane. We also show the isochrone from a polluted rMS population
scenario and how it overlaps the bMS isochrone at brighter magnitudes
whilst merging with the rMS isochrone at fainter magnitudes. This
behavior is described in more detail in Section 3.2 below.

Only certain combinations of the amount and helium fraction
of the accreted material will shift isochrones to overlap the bMS.
In Figure \ref{fig:deltat} we plot temperature shift produced by
various amounts of accreted mass.  The several curves display the
shift as a function of the helium fraction ($Y$) of the accreted
material.  The horizontal axis is defined as the temperature
difference between the bMS isochrone and that of the accreted
population after a certain mass was added. It makes little
difference if either rMS or MintP stars are the ones being
polluted, since they start with similar masses and temperatures.
The only allowed models are those that lie along $\Delta T = 0$.
Only a very limited combination of parameters can provide the
necessary shift of the isochrones required to produce the bMS
population.  The amount required is insensitive to the time at
which the accretion takes place:  later times increased
$\Delta T$ by only $\sim 26{\rm K}$ per 3 Gyr time increment.
No amount of material with less than 45\% helium would be adequate
based upon extrapolation (we did not explore accretion with greater than 20\% of the
original starting mass). If the pollution occurred more recently
than the ZAMS point, in the last few Gyr, then pollution with 
$\sim 35\%$ helium would work.

In Table 4 we show the two of our pollution scenarios that
produced the bMS population 13 Gyr isochrone, along with the TRGB
mass and helium core mass of these models. In both of these
scenarios the stars being polluted were rMS models. Results for
the pollution of MintP models were very similar, requiring only slightly
more mass in accreted material. The effects of
polluting the stars of the rMS with the parameters do not
appreciably affect the mass of the stars at the TRGB and that the
mass of the helium core at the TRGB hardly changes. A ZAHB
population generated with these stars as progenitors should be
very similar to one created by regular rMS stars or MintP stars of
normal helium content.

In Figure 9 we show the ZAHBs produced by polluted rMS and
polluted MintP stars as well as those of the rMS, MintP and bMS
stars. We also show, as thicker lines, the mass loss we inferred for
the rMS stars from the CMD of $\omega$ Centauri, the same mass
loss applied to the MintP and the polluted rMS and polluted MintP
models as well as the equivalent mass loss of the bMS stars. The
polluted-rMS HB is very similar to that of the rMS HB itself with
the stars populating a similar range of color and only some 0.25
magnitudes more luminous (vs.\ 0.8 magnitudes for helium-rich
homogeneous bMS stars) due to a combination of the slight
metallicity enhancement and helium enhancement due to the
pollution. A similar result is seen for the MintP and polluted
MintP HB though with the stars being shifted in color to the red
compared to the rMS stars by some 0.3 magnitudes in \bv. We find
that these polluted stellar models when ascending the giant branches
develop deep outer convective zones which result in the material accreted
earlier being mixed with the original stellar material and thus
being diluted by a factor of $\sim 5$. Therefore, the metallicity
and helium content at the TRGB will be close to that of the
original unpolluted star as seen in Table 4, where we list the ZAHB
parameters for the polluted scenarios which can be contrasted with
the parameters for the parent stars listed in Table 1. Thus the
polluted star descendants will arrive on the ZAHB in a similar
location on the CMD for a given mass loss, to be contrasted with
the results derived earlier for bMS stars born with a homogeneous
high level of helium.

Using the data in Table 4, we calculated the amount of helium
required to pollute the rMS stars to the required level as
compared to the amount of helium these stars would contain if
their helium levels were homogeneous as for the bMS stars. Taking
a typical $0.6 M_{\sun}$ main sequence star, we see that the mass
of helium in the accreting material varies from $0.022 \-- 0.034
M_{\sun}$ depending upon the helium content of the polluting
material. Now this same star if fully made up of helium rich
material ($Y = 0.38$) would contain some $0.228 M_{\sun}$ of helium.
We find that the polluted stars require approximately
$0.1 - 0.15$ times the amount of helium compared to the
homogeneous high helium content stars.

We calculated the amount of helium that can be produced from the
rMS stars that have already ended their lives, assuming such a
population directly produced the observed bMS population. We assumed
a total stellar mass of $5 \times 10^{6}$ solar masses for
$\omega$ Centauri today. We took yield calculations from the 
literature \citep{por98,van97,ven05}. Using a variety of IMF's 
including the Salpeter IMF \citep{sal55}, the
Miller-Scalo IMF \citep{mil79}, the Kroupa-Tout-Gilmore IMF \citep{kro93},
the Scalo IMF \citep{sca98} and the Kroupa IMF \citep{kro01} we
calculated the mass of helium produced and ejected back into
the cluster environment assuming none was lost from the cluster.
We took account of the evolutionary lifetimes of the rMS and
bMS stars as represented by the different RGB tip masses at our
assumed age of $\omega$ Centauri of 13 Gyrs. This allowed us to calculate
the mass of the original rMS and bMS populations at their formation from
the current observed number fractions of these populations. We also calculated
the mass of helium required to create the bMS as a primordial helium rich
population. Our results are shown in Table 5. We note that \citet{tsu07}
perform a similar calculation for the Salpeter IMF and show results
similar to ours.

We find that, depending upon the IMF chosen, that no more than 25$\%$
of the helium required to make a helium rich bMS population can be
produced from the intermediate mass AGB stars ($3 - 8$ solar masses)
of the rMS population. For the Kroupa-Tout-Gilmore IMF this number 
is about 1$\%$. Even if all the stars above the current RGB tip
mass are included in the calculation the mass of helium is still
only from $20\% - 75\%$ of that required and this material would
be of a far lower helium fraction than that incorporated in the
proposed levels for the bMS stars. We thus find it implausible that
the bMS population can be born from the helium rich AGB ejecta of the
rMS population.

We note that the pollution scenarios require
only some $10\% - 15\%$ of helium compared to the homogeneous
bMS stars. Such helium amounts are within the bounds of all
but the aforementioned Kroupa-Tout-Gilmore IMF choice.
The calculations above do not vary appreciably for alternative
yields in the literature for helium production. \citet{kar06}
point out that another problem for the rMS AGB stars to have
formed the bMS stars is that for AGB helium yields of the
necessary magnitude ($Y > 0.3$) the total C+N+O levels would be
increased enormously. This is inconsistent with the observed
constant values of the C+N+O abundances.

\subsection{Observational consequences of pollution models}

An interesting difference between the polluted model
isochrone and our theoretical bMS isochrone is shown
in Figure \ref{fig:modelseq}. Both sets of models
exhibit a gradual merging with the rMS isochrone but
at a significantly different bolometric luminosity. It is seen
in \citet{vil07} that the observed bMS and rMS sequences appear
to merge at magnitude of $R \sim 21.3$. We find with our models
that this merging corresponds to a $M_{\rm bol} \sim 7.69$
for bMS stars with a mass of $0.410M_{\sun}$ and 
$M_{\rm bol} \sim 7.65$ for rMS stars with a mass of $0.495M_{\sun}$.
However, we find that our bMS and rMS isochrones do not merge until
$M_{\rm bol} \sim 8.8$ some 1.1 magnitudes fainter than is observed. In
contrast our pollution scenario 2 from Table 4 merges with the rMS
isochrone at $M_{\rm bol} \sim 7.5$. We note that the
luminosity at which the merging takes place with stellar
models of $0.45M_{\sun}$ or less is sensitive primarily to the
choice of the equation of state employed in the stellar model
calculations. \citet{sol07} showed theoretical models calculated
using the evolution code of \citet{str97} using a different equation
of state than ours and that their bMS and rMS sequences merged close
to the observed merge point. We also tested our models with the old Yale
equation of state and found that the merging did not occur until
$M_{\rm bol} \sim 10$.

Using our theoretical models we calculated the expected
$(B-R)/(B+V+R)$ statistic for a composite population of the bMS
($30\%$), rMS ($65\%$) and the known metal rich population (${\rm
[M/H]} \sim -0.6$) numbering some $5\%$ of the stars. We
calculated this statistic for three mean mass losses for the rMS
population and for bMS stars coeval, 2 Gyr and 4 Gyr younger than
the majority rMS population. We repeated this for polluted rMS
models representing the bMS stars fraction of the composite
population. The results of this are plotted in Figure 10 where the
labeled cases (A thru F) are described in Table 5.
Using the observed CMD, we calculated the $\omega$ Centauri
$(B-R)/(B+V+R)$ statistic to be $0.85 \pm 0.01$.
We predict various age differences for the rMS and bMS populations
that both depend on the helium content and mass loss. These age
differences are shown in Table 6. We can see immediately that for
a given mass loss and $(B-R)/(B+V+R)$ statistic value that the age
differences for the composite population containing bMS stars
implies a much larger age difference than if the bMS stars are
replaced by polluted rMS stars.

The constraint on the age difference between the various
populations of $\omega$ Centauri has been previously investigated.
Using the morphology of the turn-off region \citet{sol05} put a
maximum on the age difference between the rMS and bMS stars
(irrespective of helium enhancement) of $\sim$ 1.5 Gyr. In their
main sequence fitting of model isochrones \citet{lee05} determined
that the age difference between the rMS and bMS stars was $\sim$ 1
Gyr. We can see from the data in Table 5 that our models predict
that only mean mass losses in the 0.2M$_{\sun}$ range with
polluted rMS stars reproduce a $(B-R)/(B+V+R)$ statistic in line
with observation. All the scenarios with the high helium bMS stars
imply age differences that are much too large. Furthermore, we
earlier calculated the mass loss range from the $\omega$ Centauri
CMD using the prominent HB clump as representing the majority rMS
population and obtained mass losses in the range $0.168
- 0.247M_{\sun}$, such mass losses are within or close to the allowable
range based upon figure 10.

\section{Summary}

It has been hypothesized that the double main sequence of $\omega$ Centauri reveals
an intermediate metal-poor population that have a large enhancement of helium, contrary to expectation of standard stellar theory. This hypothesis seemed
even more credible after spectroscopic analysis confirmed that the blue main sequence stars were indeed more metal-rich than the majority red main sequence. The proposed helium levels in these stars
are extremely large compared to known stellar populations, imply an extremely large and preferential enrichment history for their formation whilst coexisting with, or forming from,
a majority stellar population that is not unusual in any noticeable way.
 
We confirmed using our latest YREC stellar models that the proposed helium rich bMS stars do lie to the blue side of the more metal-poor rMS stars.
We developed a new procedure to allow mass loss comparison between stars of different
TRGB masses so as to predict their HB morphology and compare to the observations. 
Under the assumption that the large clump of the $\omega$ Centauri HB is a product of the
majority rMS metal-poor population, we inferred that the mass loss suffered by these stars on the RGB spans a range of $0.168 - 0.247M_{\sun}$ from TRGB progenitors
of $0.814M_{\sun}$ at an age of 13 Gyrs. This in turn implied that the proposed high helium bMS stars would have lost a minimum of $0.102 - 0.147M_{\sun}$ from a TRGB progenitor
of some $0.629M_{\sun}$ at the same age. We find such helium rich stars would appear on the HB in the blue tail below the majority clump that is assumed to be from the rMS stars. If we assume all
such stars on the CMD of the $\omega$ Centauri HB are attributable to the bMS population then we have only $\sim 25\%$ of the bMS plus rMS stars residing there. This seems distinctly
at odds with the $\sim 37\%$ of bMS plus rMS star total we observe in the MS itself. We also note that if we use our Case II or Case III mass losses for these stars the HB bMS stars should
appear far down the blue tail in a large clump clearly separated from the observed majority clump we identify with the rMS stars. Such a disconnected clump at the bottom of the
blue tail of the $\omega$ Centauri HB is not seen at all.

Unless the mass loss mechanism for the proposed high helium bMS stars operates differently from the rMS stars causing them the reside in the large observed
clump then these stars are absent from the HB of $\omega$ Centauri. In fact, for the bMS stars to be hiding amongst the majority HB clump then their mass loss must be
in the range $0.059 - 0.101M_{\sun}$, where the lower limit is determined by the fact we do not see overluminous HB stars in substantial numbers above the main HB clump.
Prior observations of the instability strip RR Lyrae variables have not found evidence of high helium stars
there, implying that the mass loss of these stars on the RGB was not far lower than what we have assumed. We noted earlier that if the bMS stars were $2 - 5$ Gyr younger they would
also appear in the instability strip of $\omega$ Centauri but such age differences have been ruled out by turn-off morphology in other studies.

It is possible to create a blue main sequence theoretically by the pollution of existing rMS or MintP stars with helium rich material for limited combinations
of helium content and mass accreted. The helium content of such material is not far from the range of the latest intermediate mass AGB stellar yields \citep{her04}, especially
if the accretion is relatively recent. The total mass of the accreted material is in line with estimates of the 
possible pollution in a globular cluster environment \citep{tho02}. We do note that the AGB helium yields we require are on the upper end (or slightly beyond)
of such calculations and that the mass accreted to create our blue main sequences is incompatible with the current $\omega$ Centauri environment, though possibly not that of past
when the cluster was not only more massive but the velocity dispersion lower; both of which would increase the amount of accretion possible onto existing stars.
Creating bMS stars out of homogeneous helium rich material is not the only possible mechanism for their formation, and pollution via accretion
of AGB ejecta in a globular cluster environment is another possibility. We see that of the observed $\omega$ Centauri RR Lyrae variables the presence
of both MintP stars and rMS stars but no helium-rich bMS stars, this implies that at least some MintP stars formed with normal helium enrichments. An accretion pollution
scenario has also been hypothesized in the past by \citet{can98} in regard to the main sequence abundance anomalies in the globular cluster 47 Tuc, where the C and N
abundance variations are seen on the main sequence and in red giants. We note our simple pollution scenario does not explain the abundance variations seen on both
the main sequence and giant branches, as they would be washed out as the stars evolve up the RGB.

The polluted star scenario predicts that the ZAHB for such a population should not differ appreciably from that of the unpolluted parent stars. We see from this that
we would observe in the $\omega$ Centauri CMD a double main sequence and a HB that possesses a majority single clump, very few over luminous stars in the horizontal region
of the HB, no large disconnected clump of stars in the HB blue tail and no RR Lyrae variables that are over luminous. This is what is observed in contrast to the predictions of the homogeneously
helium rich bMS stars that have been proposed. The polluted star scenario though requiring certain combinations of pollution helium content and total mass accreted does not
require the very large helium amounts that the proposed bMS stars do, amounts that seem greater than that possible from ejecta of more massive rMS stars in $\omega$ Centauri.

Another prediction from our modeling is that the polluted stars will merge with the rMS stars at approximately the observed luminosity whereas our model homogeneous bMS stars do not
do this until about 1.1 magnitudes fainter. The position of the this merger may turn out to be an important diagnostic in determining the composition of the bMS stars.

We used the $(B-R)/(B+V+R)$ statistic that is usually applied to the investigation of single stellar population horizontal branches in globular clusters. When we apply
our calculated $(B-R)/(B+V+R)$ of $0.85 \pm 0.01$ for $\omega$ Centauri to theoretical predictions of this statistic using our models we found that the high helium bMS star models
predict age differences for the bMS population that are too great when compared to turn-off morphology estimates. The pollution scenario for mass losses less than $\sim 0.2M_{\sun}$ are
within the allowable limit of age differences. A result that is also approximately the range of estimated mass losses for the rMS population ($0.168 - 0.247M_{\sun}$)
we inferred from the $\omega$ Centauri CMD using our theoretical models.


\acknowledgements
We would like to thank M. Pinsonneault and D. Weinberg for helpful suggestions and A. Sollima for kindly furnishing to us the HST data. We 
would also like to thank the anonymous referee whose many and detailed comments have greatly improved this paper.

\clearpage

\begin{deluxetable}{lccc}
  \tablecaption{Model population parameters.\label{parameters}}
  \tablewidth{0pt}
  
  \tablenum{1}
  
  \tablehead{
   \colhead{Parameter} &
   \colhead{rMS} &
   \colhead{bMS} &
   \colhead{MintP}
  }
\startdata
 $Z$                   & 0.0005 & 0.001  & 0.001  \\
 $Y$                   & 0.231  & 0.382  & 0.232  \\
 ${\rm [M/H]}$         & --1.57 & --1.27 & --1.27 \\
 TRGB mass ($M_\odot$) & 0.814  & 0.629  & 0.826  \\
 TRGB core mass ($M_\odot$) & 0.485  & 0.455  & 0.479  \\
 \enddata
\end{deluxetable}

\clearpage

\begin{deluxetable}{cccc}
 \tablecaption{Mass loss for the bMS population ($M_\odot$)\label{deltam}.}
 \tablewidth{0pt}
 
 \tablenum{2}
 
 \tablehead{
  & \multicolumn{3}{c}{bMS mass loss}\\ \cline{2-4}
  \colhead{rMS Mass loss}
  & \colhead{Case I}
  & \colhead{Case II}
  & \colhead{Case III}
  }
 \startdata
0.12 & 0.072 & 0.084 & 0.114 \\
0.16 & 0.097 & 0.112 & 0.152 \\
0.20 & 0.118 & 0.140 & 0.190 \\
0.24 & 0.143 & 0.168 & 0.228 \\
0.28 & 0.168 & 0.196 & 0.266 \\
\enddata
\end{deluxetable}

\clearpage

\begin{deluxetable}{cccccc}
 \tablecaption{Pollution scenario parameter matrix.}
 \tablewidth{0pt}
 
 \tablenum{3}
 
 \tablehead{
  \colhead{Population} & \colhead{Fractional Mass} & \colhead{X} & \colhead{Y} & \colhead{Z} & \colhead{Accretion Age} \\
\colhead{} & \colhead{} & \colhead{} & \colhead{} & \colhead{} &
\colhead{(Gyr)} }

\startdata
MintP & 0.2 & 0.699 & 0.30 & 0.001 & 0.0 \\
rMS & 0.159 & 0.649 & 0.35 &  & 3.0 \\
 & 0.129 & 0.599 & 0.40 &  & 6.0 \\
 & 0.1 & 0.549 & 0.45 &  & 9.0 \\
 & 0.07 & 0.499 & 0.50 &  &  \\
 & 0.04 &  &  &  &  \\
 \\
 \\
\enddata

\end{deluxetable}

\clearpage

\begin{deluxetable}{cccccc}

\tablecaption{Pollution scenarios and their calculated ZAHB
properties.}
 \tablewidth{0pt}
 
 \tablenum{4}

\tablehead{\colhead{Population} & \colhead{Fractional Mass} & \colhead{Z} & \colhead{Y} & \colhead{TRGB Mass} & \colhead{Helium Core Mass} \\
\colhead{} & \colhead{} & \colhead{} & \colhead{} &
\colhead{($M_{\sun}$)} & \colhead{($M_{\sun}$)} }

\startdata
Scenario 1 & 0.07 & 0.001 & 0.5 & 0.810 & 0.481 \\
Scenario 2 & 0.129 & 0.001 & 0.45 & 0.811 & 0.481 \\
 &  &  &  &  \\
Polluted rMS ZAHB & - & 0.0006 & 0.276 & - & 0.481 \\
Polluted MintP ZAHB & - & 0.001 & 0.276 & - & 0.481 \\

\enddata

\end{deluxetable}

\clearpage

\begin{deluxetable}{cccc}

\tablecaption{Helium Yield Calculations for different IMF's}
 \tablewidth{0pt}

\tablenum{5}

\tablehead{\colhead{IMF} & \colhead{Helium required} & \colhead{Helium produced} & \colhead{Helium produced} \\
\colhead{} & \colhead{($M_{\sun}$)} & \colhead{(all stars)($M_{\sun}$)} &
\colhead{(AGB stars only)($M_{\sun}$)} } \startdata
Salpeter & $5.1 \times 10^{5}$ & $2.3 \times 10^{5}$ & $5.9 \times 10^{4}$ \\
Miller-Scalo & $1.0 \times 10^{6}$ & $6.8 \times 10^{5}$ & $2.5 \times 10^{5}$ \\
Scalo & $7.0 \times 10^{5}$ & $4.0 \times 10^{5}$ & $1.1 \times 10^{5}$ \\
Kroupa-Tout-Gilmore & $4.3 \times 10^{5}$ & $8.6 \times 10^{4}$ & $4.0 \times 10^{3}$ \\
Kroupa & $8.8 \times 10^{5}$ & $6.5 \times 10^{5}$ & $1.6 \times 10^{5}$ \\
\enddata

\end{deluxetable}

\clearpage

\begin{deluxetable}{cccc}

\tablecaption{Calculated age differences.}
 \tablewidth{0pt}

\tablenum{6}

\tablehead{\colhead{Case} & \colhead{Mean rMS mass loss} & \colhead{Helium enhancement} & \colhead{Age difference} \\
\colhead{} & \colhead{($M_{\sun}$)} & \colhead{} &
\colhead{(Gyrs)} } \startdata
A & 0.20 & Homogeneous & $-1.55$ - $-1.75$ \\
B & 0.24 & Homogeneous & $-3.35$ - $-3.80$ \\
C & 0.28 & Homogeneous & $-5.55$ - $-6.15$ \\
 \\
D & 0.20 & Polluted & $-0.65$ - $-0.75$ \\
E & 0.24 & Polluted & $-2.15$ - $-2.25$ \\
F & 0.28 & Polluted & $-2.45$ - $-2.55$ \\
\enddata

\end{deluxetable}

\clearpage

\clearpage

\begin{figure}
\epsscale{0.85} \plotone{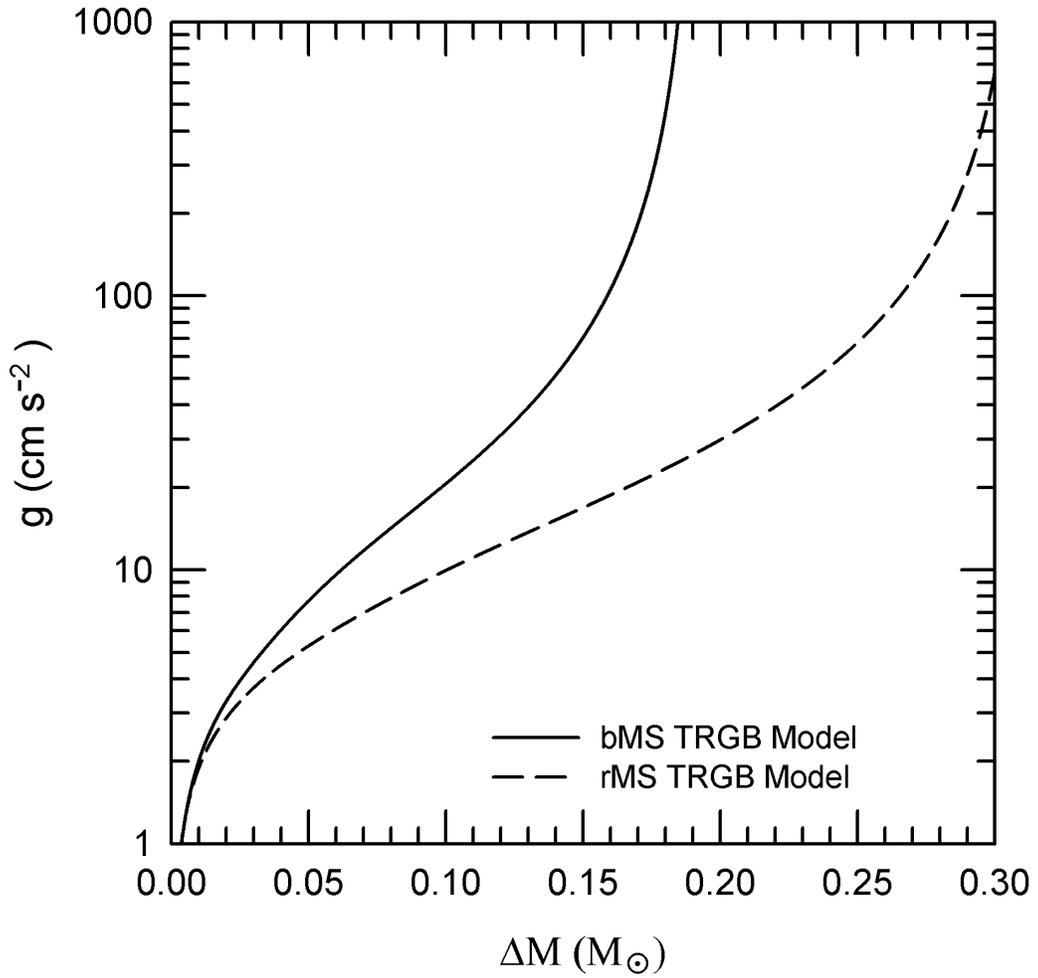} \caption{Stratification of the
rMS and bMS models near the TRGB.  The plot shows the
gravitational acceleration as a function of stellar mass below the
surface. \label{fig:deltam}}
\end{figure}
\clearpage

\begin{figure}
\epsscale{0.85} \plotone{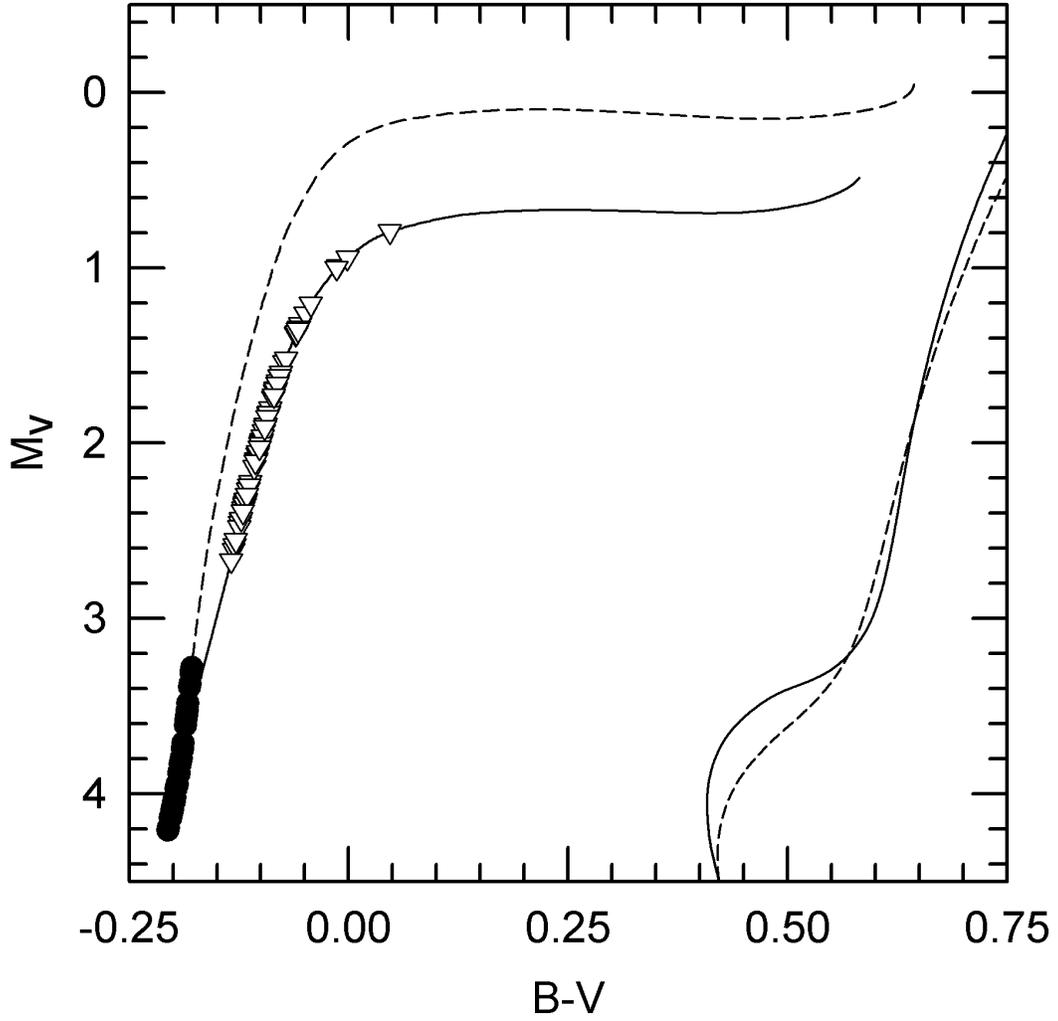} \caption{rMS (open triangles) and
bMS (filled circles) 13 Gyr ZAHBs populated for a rMS mean mass
loss of 0.24$M_{\sun}$ and a bMS mean mass loss of
0.143$M_{\sun}$ using Case I mass losses.\label{fig:hbmodels}}
\end{figure}
\clearpage

\begin{figure}
\epsscale{0.85} \plotone{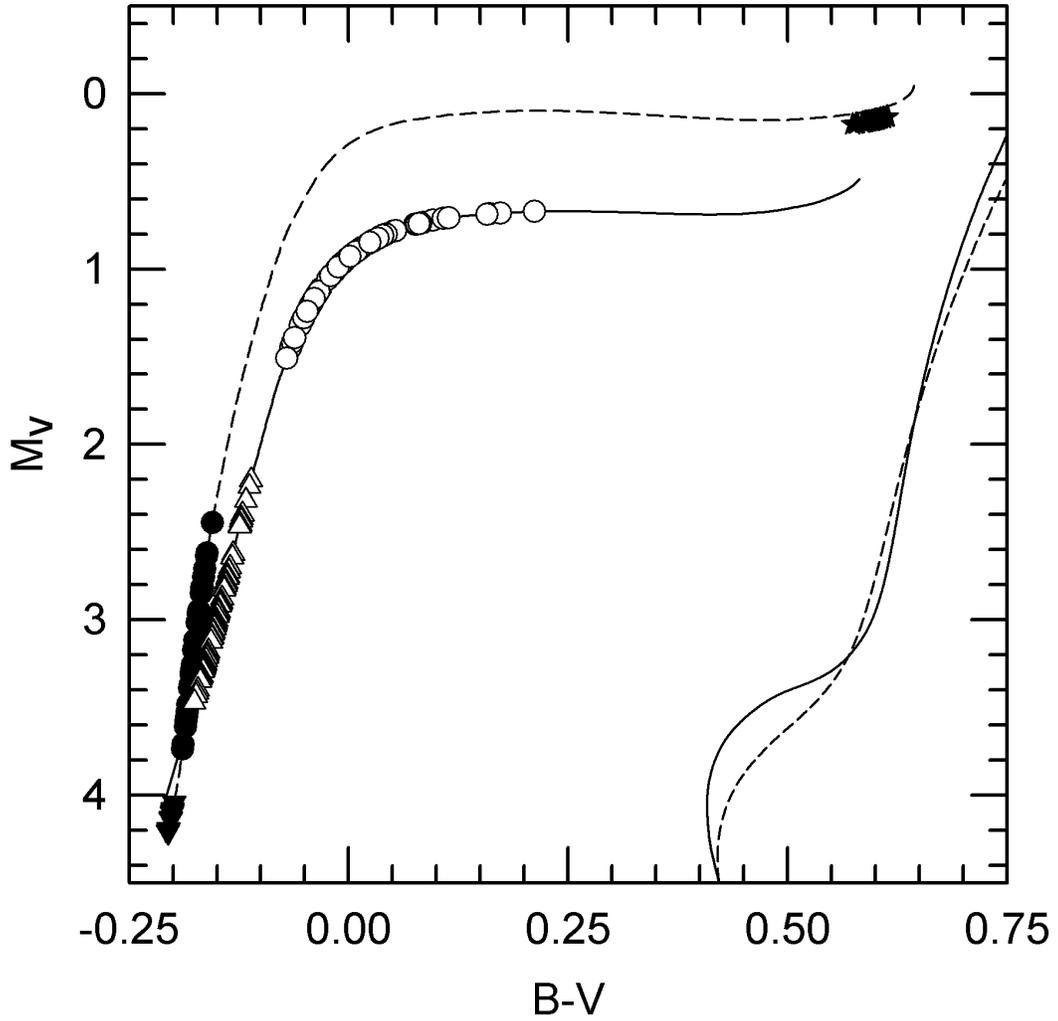} \caption{The rMS 13 Gyr ZAHBs
populated for rMS mean mass losses of 0.20$M_{\sun}$ (open
circles) and 0.28 $M_{\sun}$ (open triangles) and the bMS 13 Gyr
population populated for bMS mean mass losses of 0.118 $M_{\sun}$
(filled circles) and 0.168 $M_{\sun}$ (filled triangles) and the
bMS 7 Gyr population with mean mass loss 0.143$M_{\sun}$ (filled
stars). All bMS mass losses are using Case I.\label{fig:altdeltam}}
\end{figure}
\clearpage

\begin{figure}
\epsscale{0.85} \plotone{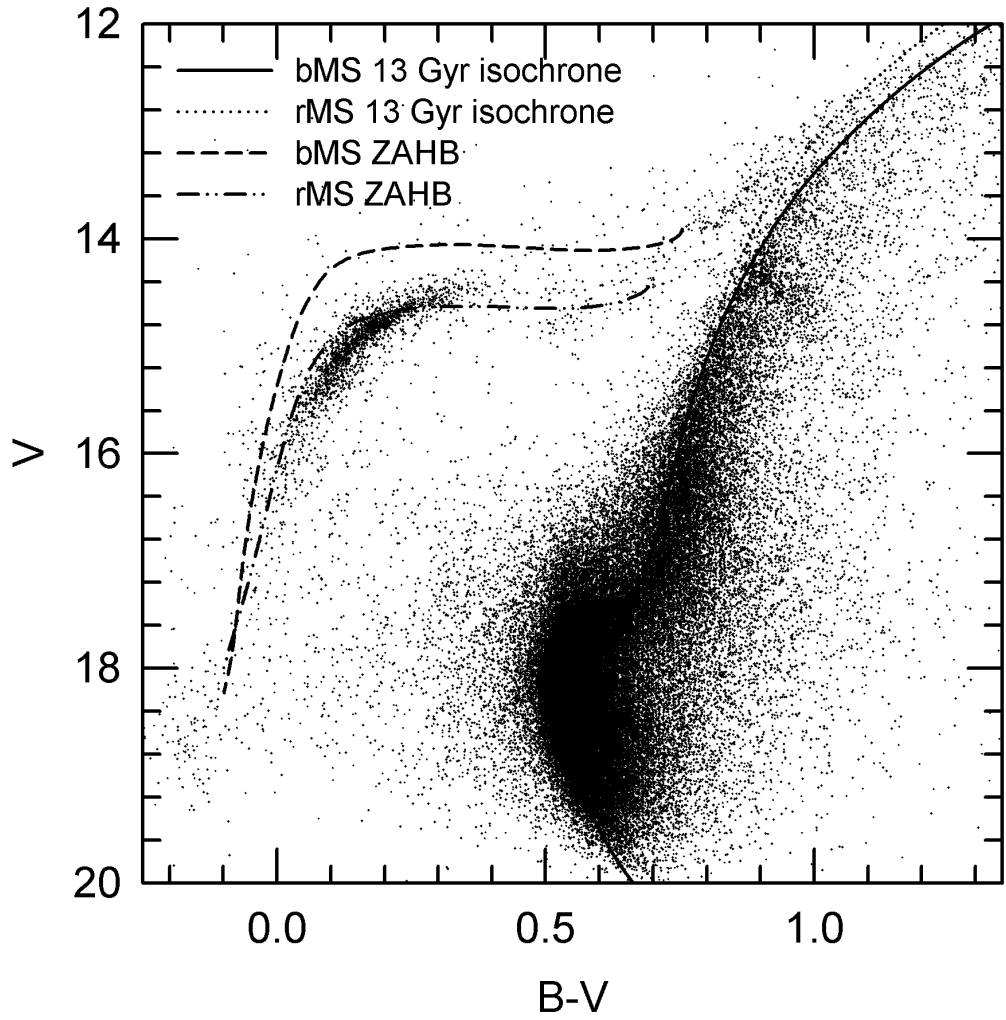} \caption{CMD of $\omega$ Centauri
with 13 Gyr rMS and bMS isochrones. Photometry is from
\citet{rey04}.\label{fig:cmd}}
\end{figure}
\clearpage

\begin{figure}
\epsscale{0.85} \plotone{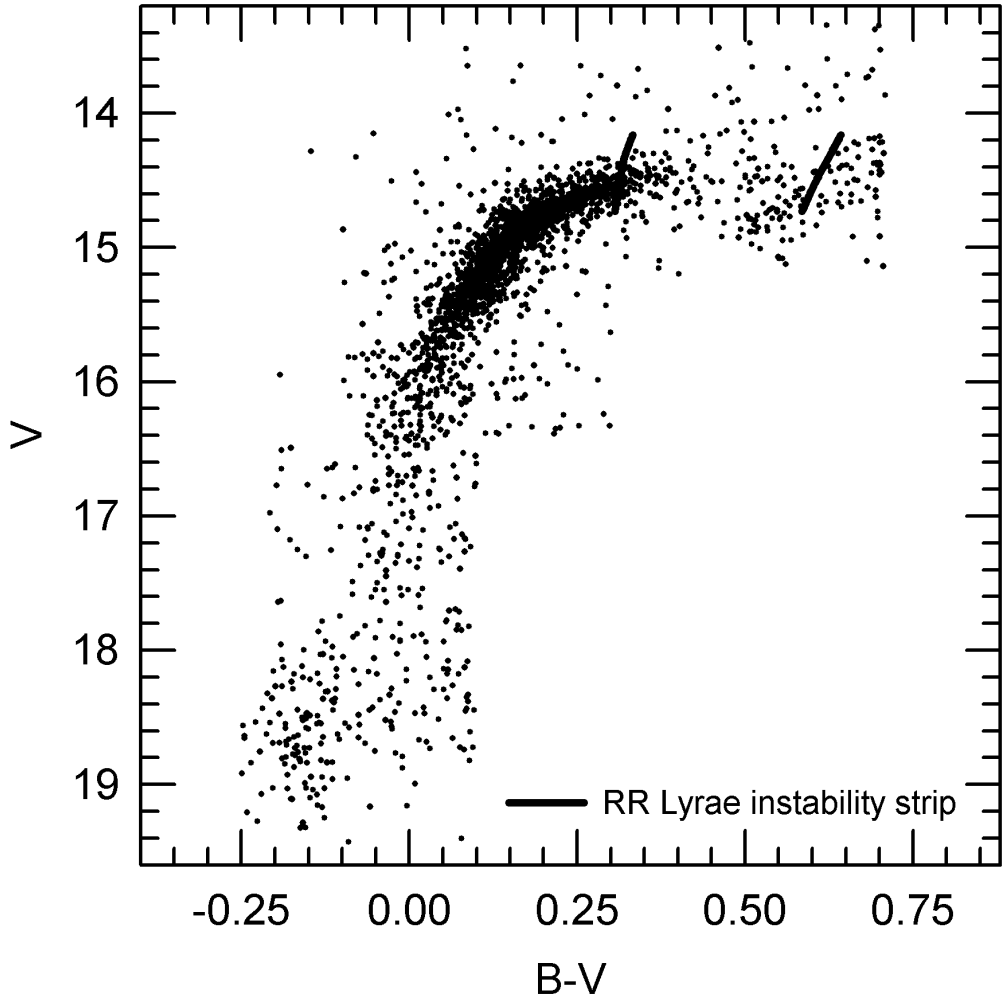} \caption{Horizontal branch CMD of
$\omega$ Centauri showing the RR Lyrae instability strip for the
rMS population.\label{fig:rrstrip}}
\end{figure}
\clearpage

\begin{figure}
\epsscale{0.85} \plotone{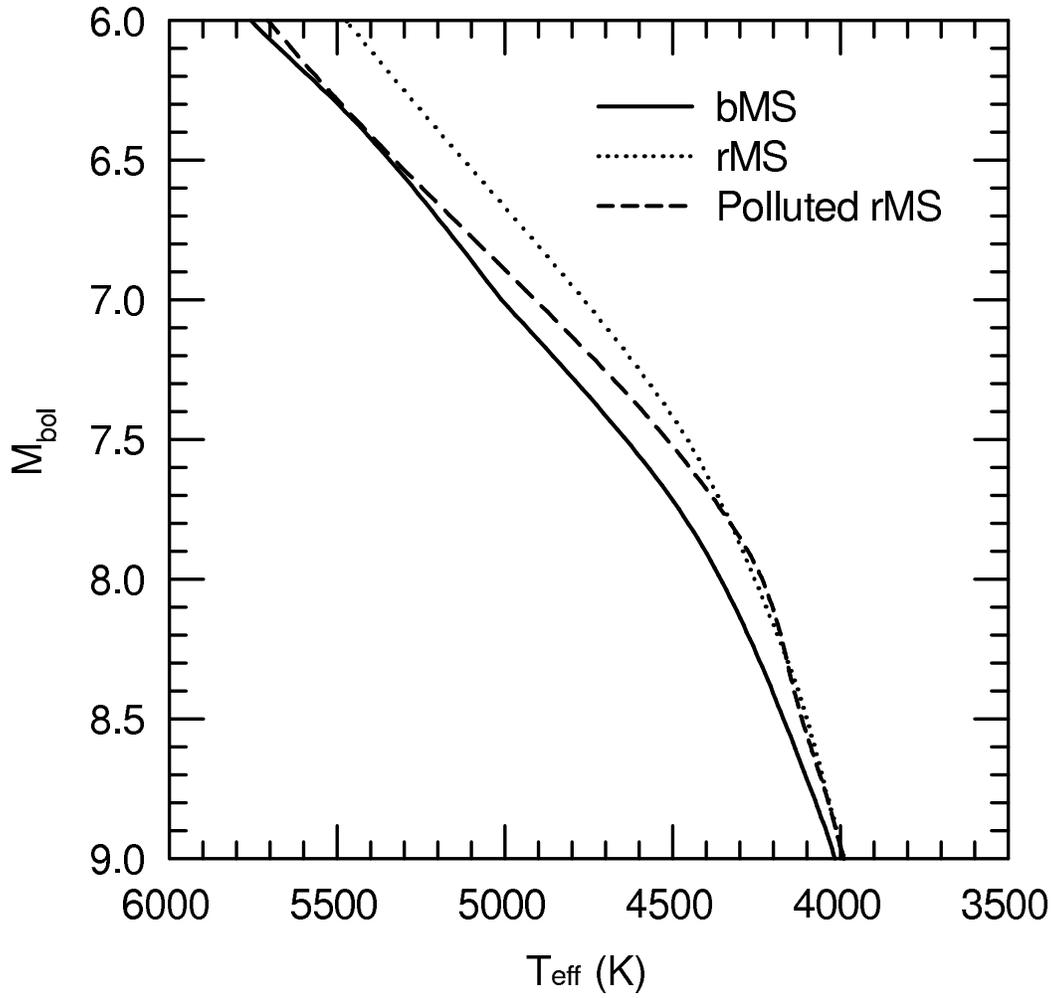} \caption{Model main sequences of
the rMS and bMS populations and a polluted rMS sequence.\label{fig:modelseq}}
\end{figure}
\clearpage

\begin{figure}
\epsscale{0.85} \plotone{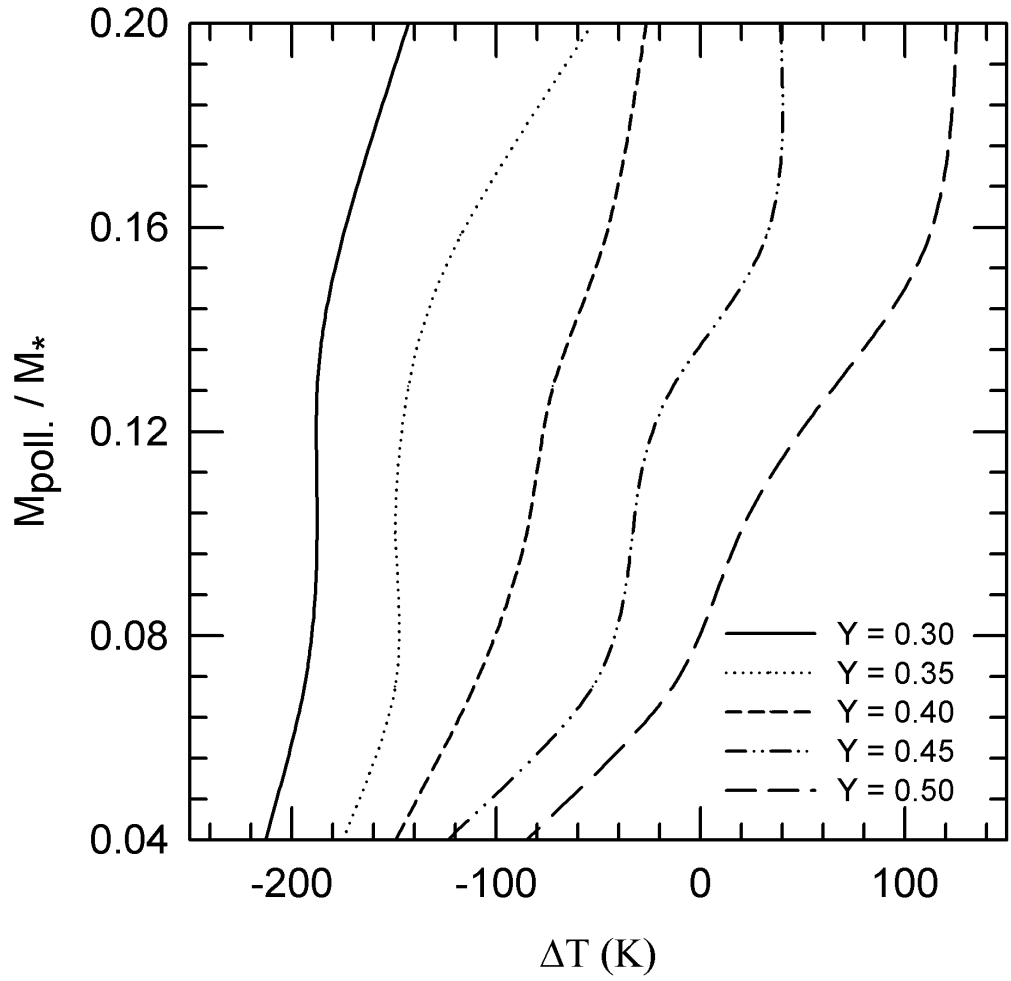} \caption{Mass fraction of pollution to
total stellar mass vs.\ change in $T_{\rm eff}$ ($\Delta$T) for rMS
stellar models.\label{fig:deltat}}
\end{figure}
\clearpage

\begin{figure}
\epsscale{0.85} \plotone{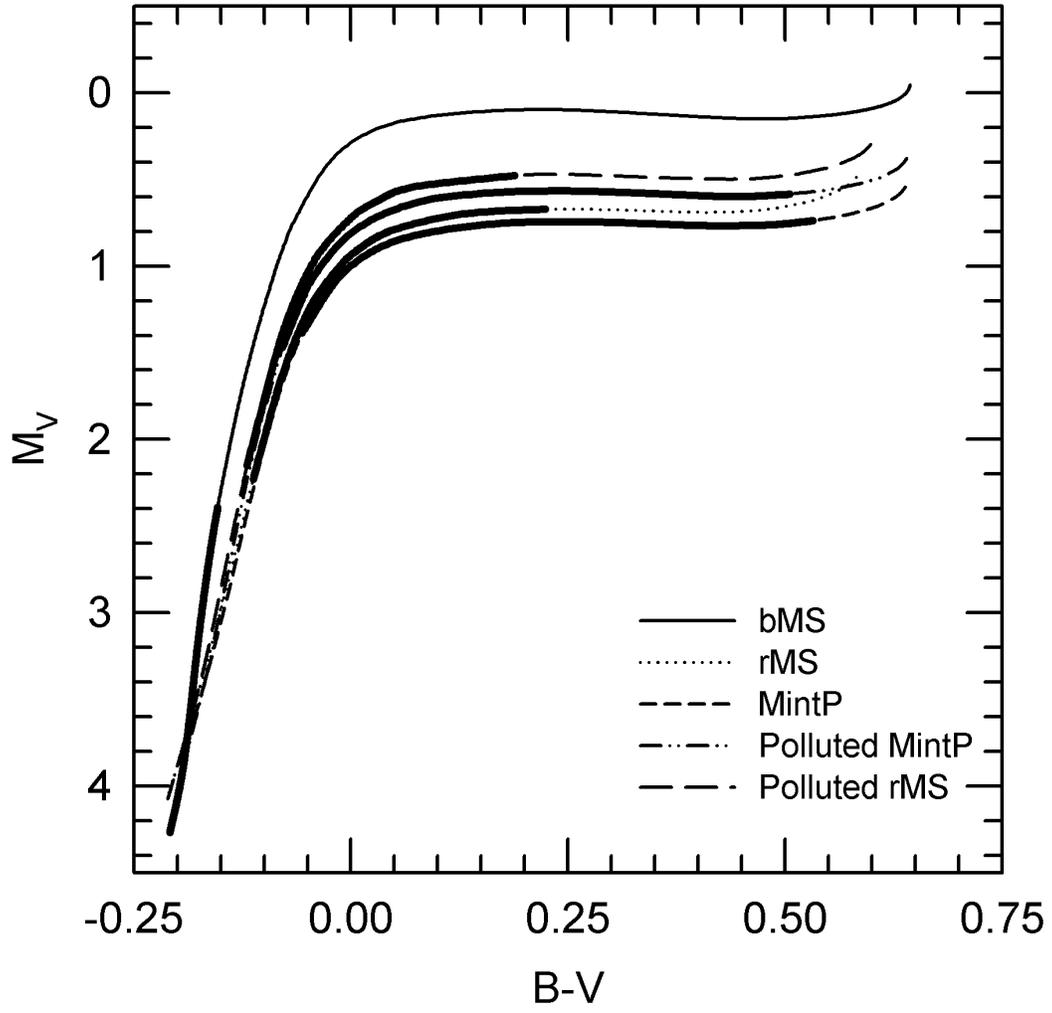} \caption{Standard and pollute
population ZAHB's and equivalent mass losses (Case I) based upon the
inferred rMS mass loss of 0.168$M_{\sun} - 0.247M_{\sun}$}
\end{figure}
\clearpage

\begin{figure}
\epsscale{0.85} \plotone{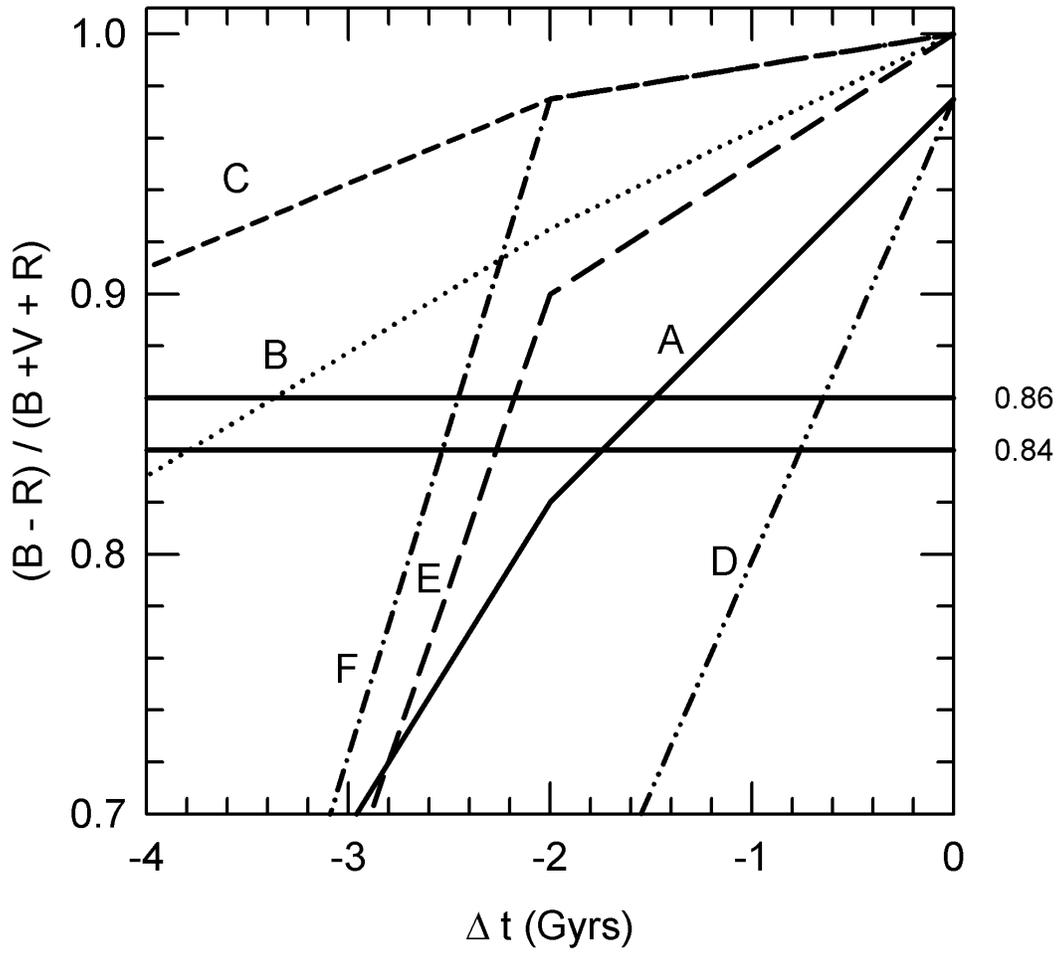} \caption{$(B-R)/(B+V+R)$
statistic vs. age difference of a composite $\omega$ Centauri
population. See Table 6 for the case (A \-- F) details.}
\end{figure}


\begin{thebibliography}

\bibitem[Alexander \& Ferguson(1994)]{ale94} Alexander, D. R.,
\& Ferguson, J. W., 1994, \apj, 437, 879

\bibitem[Allard \& Hauschildt(1995)]{all95} Allard, F., \&
Hauschildt, P. H. 1995, \apj, 445, 433

\bibitem[An et al.(2007)]{an07} An, D., Terndrup, D. M.,
Pinsonneault, M. H., Paulson, D. B., Hanson, R. B., \& Stauffer,
J. R., 2007, \apj, 655, 233

\bibitem[Bahcall \& Loeb(1990)]{bah90} Bahcall, J. N.,
\& Loeb, A., 1990, \apj, 360, 267

\bibitem[Bahcall et al.(2001)]{bah01} Bahcall, J. N.,
Pinsonneault, M. H., \& Basu, S., 2001, \apj, 555, 990

\bibitem[Bedin et al.(2004)]{bed04} Bedin L., Piotto G.,
Anderson J., Cassisi S., King I., Momany Y., \& Carraro G., 2004,
\apj, 605, L125

\bibitem[Bekki \& Norris(2006)]{bek06} Bekki, K., \& Norris,
J., 2006, \apj, 637L, 109

\bibitem[Bergbusch \& Vandenberg(1992)]{ber92} Bergbusch, P. A.,
\& Vandenberg D. A., 1992, \apjs, 81, 163

\bibitem[B\"{o}hm-Vitense(1958)]{boh58} B\"{o}hm-Vitense, E. 1958,
Z. Astrophys., 46, 108

\bibitem[Bondi(1952)]{bon52} Bondi, H., 1952, MNRAS, 112, 195

\bibitem[Bono et al.(2002)]{bono02} Bono, G., Balbi, A.,
Cassisi, S., Vittorio, N., \& Buonanno, R. 2002, \apj, 568, 463

\bibitem [Bono et al.(1995)]{bon95} Bono, G., Caputo, F.,
\& Marconi, M. 1995, \aj, 110, 2365

\bibitem[Cannon et al. (1998)]{can98} Cannon, R. D., Croke,
B. F. W., Bell, R. A., Hesser, J. E., \& Stathakis, R. A., 1998,
\mnras, 298, 601

\bibitem[Catelan(2000)]{cat00} Catelan, M., 2000, \apj, 531, 826

\bibitem[Choi \& Yi(2007)]{cho07} Choi, E., \& Yi, S. K., MNRAS, 375, L1

\bibitem[Cox et al.(1968)]{cox68} Cox, J. P., \& Guili,
R. T., 1968, Principles of Stellar Structure (New York: Gordon and
Breach)

\bibitem[D'Antona et al.(2002)]{dan02} D'Antona, F.,
Caloi, V., Montalban, J., Ventura, P., \& Gratton, R., 2002, \aap,
395, 69

\bibitem[D'Antona et al.(2005)]{dan05} D'Antona, F.,
Bellazzini, M., Caloi, V., Fusi-Pecci, F., Galleti, S., \& Rood,
R. T., 2005, \apj, 631, 868

\bibitem[D'Cruz et al.(2000)]{dcr00} D'Cruz et al., 2000,
\apj, 530, 352

\bibitem[Delahaye \& Pinsonneault(2005)]{del05} Delahaye,
F., \& Pinsonneault, M., 2005, \apj, 625, 563

\bibitem[Dickens \& Wooley(1967)]{dic67} Dickens, R. J.,
\& Wooley S. J., 1967, Royal Obs. Bull., 128, E255

\bibitem[Ferraro et al.(2004)]{fer04} Ferraro, F. R., Sollima, A., 
Pancino, E., Bellazzini, M., Origlia, L., Straniero, O., \& Cool, A.,
2004, \apj, 603, L81

\bibitem[Freeman \& Rodgers(1975)]{fre75} Freeman, K. C.,
\& Rodgers, A. W., 1975, \apj, 201, L71

\bibitem[Fukugita \& Kawasaki(2006)]{fk06} Fukugita, M., \&
Kawasaki, M. 2006, \apj, 646, 691

\bibitem[Green et al.(1987)]{gre87} Green, E. M., Demarque, P.,
\& King, C. R., 1987, the Revised Yale Isochrones and Luminosity
Functions (New Haven: Yale Univ. Obs.)

\bibitem[Grevesse \& Noels(1993)]{gre93} Grevesse, N., \& Noels,
A., 1993, in Origin and Evolution of the Elements, ed. N.
Prantzos, E. Vangioni-Flam, \& M. Cassé (Cambridge: Cambridge
Univ. Press), 15

\bibitem[Gruzinov \& Bahcall(1998)]{gru98} Gruzinov, A., \&
Bahcall, J., 1998, \apj, 504, 996

\bibitem[Guenther et al.(1992)]{gue92} Guenther, D. B.,
Demarque, P., Kim, Y.-C., \& Pinsonneault, M. H., 1992, \apj, 387,
372

\bibitem[Herwig (2004)]{her04} Herwig, F., 2004, \apjs, 155, 651

\bibitem[Iglesias \& Rogers(1996)]{igl96} Iglesias, C. A.,
\& Rogers, F. J., 1996, \apj, 464, 943

\bibitem[Itoh et al.(1996)]{ito96} Itoh, N., Hayashi, H., \&
Nishikawa, A., 1996, \apjs, 102, 411

\bibitem[Jimenez et al.(2003)]{jim03} Jimenez, R., Flynn, C.,
MacDonald, J., \& Gibson, B. K., 2003, Science, 299, 1552

\bibitem[Karakas et al.(2006)]{kar06} Karakas, A., Fenner, Y.,
Sills, A., Campbell, S. W., \& Lattanzio, J. C., 2006, preprint
(astro-ph/0608366)

\bibitem[Kroupa et al.(1993)]{kro93} Kroupa, P., Tout, C. A., \&
Gilmore, G., 1993, MNRAS, 262, 545

\bibitem[Kroupa (2001)]{kro01} Kroupa, P., 2001, MNRAS, 322, 231

\bibitem[Lee et al.(1994)]{lee94} Lee Y.-W., Demarque
P., \& Zinn R., 1994, \apj, 423, 248

\bibitem[Lee et al.(2005)]{lee05} Lee Y.-W., Joo S-J.,
Han S-I., Chung C., Ree C., Sohn Y-J., Kim Y-C., Yoon S-J., Yi S.
\& Demarque P., 2005, \apj 621, L57

\bibitem[Maeder \& Meynet(2006)]{mae06} Maeder, A., \&
Meynet, G., 2006, \aap, 448, L37

\bibitem[Miller \& Scalo(1979)]{mil79} Miller G. E., \&
Scalo, J. M., 1979, \apjs, 41, 513

\bibitem[Momany et al.(2004)]{mom04} Momany, Y., Bedin, L. R.,
Cassisi, S., Piotto, G., Ortolani, S., Recio Blanco, A., De
Angeli, F., \& Castelli, F., 2004, \aap, 420, 605

\bibitem[Norris(2004)]{nor04} Norris, J. E., 2004,
\apj, 612, L25

\bibitem[Norris \& Da Costa(1995)]{nor95} Norris, J.,
\& De Costa G. S., 1995, \apj, 447, 680

\bibitem[Norris et al.(1996)]{nor96} Norris, J. E.,
Freeman, K. C., \& Mighell, K. J., 1996, \apj, 462, 241

\bibitem[Olive \& Skillman(2004)]{oli04} Olive, K. A., \&
Skillman, E. D. 2004, \apj, 617, 29

\bibitem[Origlia et al.(2002)]{ori02} Origlia, L., Ferraro,
F. R., Fusi Pecci, F., \& Rood, R. T., 2002, \apj, 571, 458

\bibitem[Pagel(1992)]{pag92} Pagel, B. E. J., 1992, IAU
Symp. 149, Stellar Populations in Galaxies, ed. B. Barbury, \& A.
Renzini, (Dordrecht: Kluwer), 133

\bibitem[Pancino et al.(2000)]{pan00} Pancino E., Ferraro F.,
Bellazzini M., Piotto G., \& Zoccali M., 2000, \apj, 534, L83

\bibitem[Pinsonneault et al.(2003)]{pin03} Pinsonneault,
M. H., Terndrup, D. M., Hanson, R. B., \& Stauffer, J. R., 2003,
\apj, 598, 588

\bibitem[Piotto et al.(2005)]{pio05} Piotto et al., 2005,
\apj, 621, 777

\bibitem[Portinari et al.(1998)]{por98} Portinari, L.,
Chiosi, C., \& Bressan, A., 1998, \aap, 334, 505

\bibitem[Rey et al.(2004)]{rey04} Rey, S.-C., Lee, Y.-W.,
Ree, C.-H., Joo, J.-M., \& Sohn Y.-J., 2004, \aj, 127, 958

\bibitem[Rogers et al.(1996)]{rog96} Rogers, F. J., Swenson,
F. J., \& Iglesias, C. A., 1996, \apj, 456, 902

\bibitem[Salpeter(1955)]{sal55} Salpeter, E., 1955, \apj, 121, 161

\bibitem [Scalo(1998)]{sca98} Scalo, J., 1998, in ASP Conf. Ser., 142, The
Stellar Initial Mass Function, 201

\bibitem[Saumon et al.(1995)]{sau95} Saumon, D., Chabrier, G.,
\& Van Horn, H. M., 1995, \apjs, 99, 713

\bibitem[Serenelli \& Weiss(2005)]{ser05} Serenelli, A., \& Weiss,
A. 2005, \aap, 442, 1041

\bibitem[Sills et al.(2000)]{sil00} Sills, A., Pinsonneault,
M. H., \& Terndrup, D. M., 2000, \apj, 534, 335

\bibitem [Sollima et al.(2007)]{sol07} Sollima, A., Ferraro, F. R.,
Bellazzini, M., Origlia, L., Straniero, O., \& Pancino, E., 2007, ApJ, 654, 915

\bibitem[Sollima et al.(2006)]{sol06} Sollima, A., Borrisova,
J., Catelan, M., Smith, H. A., Minniti, D., Cacciari, C., \&
Ferraro, F. R., 2006, \apj, 640, L43

\bibitem[Sollima et al.(2005)]{sol05} Sollima, A., Pancino,
E., Ferraro, F., Bellazzini, M., Straneiro, O., \& Pasquini, L.,
2005, \apj, 634, 332

\bibitem [Straniero et al.(1997)]{str97} Straniero, O., 
Chieffi, A. \& Limongi, M., 1997, ApJ, 490, 425

\bibitem[Suntzeff \& Kraft(1996)]{sun96} Suntzeff, N. B.,
\& Kraft, R. P., 1996, \aj, 111, 1913

\bibitem[Thoul et al.(1994)]{tho94} Thoul, A. A., Bahcall,
J. N., \& Loeb, A., 1994, \apj, 421, 828

\bibitem[Thoul et al.(2002)]{tho02} Thoul, A., Jorissen, A.,
Goriely, S., Jehin, E., Magain, P., Noels, A., \& Parmantier, G.,
2002, \aap, 383, 491

\bibitem[Thuan \& Izotov(2002)]{thuan02} Thuan, T. X., \&
Izotov, Y. I. 2002, Space Sci.\ Rev., 100, 263

\bibitem[Tsujimoto et al.(2007)]{tsu07} Tsujimoto, T., Shigeyama, T.,
\& Suda, T., 2007, \apj, 654, L139

\bibitem[Vandenberg \& Clem(2003)]{van03} VandenBerg, D. A.,
\& Clem, J. L., 2003, \aj, 126, 778

\bibitem[Van den Hoek \& Groenewegen(1997)]{van97} Van den
Hoek, L. B., \& Groenewegen, M. A. T., 1997, \aaps, 123, 305

\bibitem[Ventura \& D'Antona(2005)]{ven05} Ventura, P., \&
D'Antona, F., 2005, \aap, 439, 1075

\bibitem [Villanova et al.(2007)]{vil07} Villanova, S., Piotto, G.,
King, I. R., Anderson, J., Bedin, L. R., Gratton, R. G., Cassisi, S.,
Momany, Y., Bellini, A., Cool, A. M., Recio-Blanco, A. \& Renzini, A.,
2007, preprint (astro-ph/0703208)

\bibitem[Weiss et al.(2006)]{wei06} Weiss, A., Salaris, M., Ferguson, J. W.,
\& Alexander, D. R., 2006, preprint (astro-ph/0605666)

\bibitem[Yi \& Demarque(2003)]{yi03} Yi, S., Kim, Y.-C.,
\& Demarque P., 2003, \apjs, 144, 259

\end{thebibliography}
\end{document}